\documentclass[10pt, conference, letterpaper]{IEEEtran}
\IEEEoverridecommandlockouts
\usepackage[T1]{fontenc}
\usepackage[english]{babel}
\usepackage[utf8]{inputenc}
\usepackage{comment,acronym,graphicx,amsthm,amsmath,amssymb,tabularx,bbm,algorithmic,subfigure,epsfig,xspace,multirow}
\usepackage[algoruled,vlined,linesnumbered]{algorithm2e}
\usepackage[dvipsnames]{xcolor}
\usepackage{flushend}
\newtheorem{lemma}{Lemma}

\newtheorem{thm}{Theorem}

\newtheorem{prob}{Problem}
\def \S{{\mathcal S}}

\newcommand\Prob[1]{{\mathbbm P} \left \{ {#1} \right \}}
\def \C{{\mathcal C}}
\def \U{{\mathcal U}}
\def \L{{\mathcal L}}
\def \bx{{\mathbf x}}
\newcommand \e[1]{{\mathbf e}_{#1}}
\def \bX{{\mathbf X}}
\newcommand \1[1]{{\mathbbm 1}\left \{{#1} \right \}}
\def \bp{{\mathbf p}}
\def \p{q}\def \P{Q}

\newcommand{\tsoff}{t_{\mbox{\tiny off}}}
\newcommand{\tson}{t_{\mbox{\tiny on}}}
\newcommand{\zoff}{z_{\mbox{\tiny off}}}
\newcommand{\zon}{z_{\mbox{\tiny on}}}

\begin{document}

\title{Optimal Control of Storage Regeneration with Repair Codes}
\author{Francesco De Pellegrini$^\diamond$, Rachid El Azouzi$^\star$, Alonso Silva$^\ddagger$ and Olfa Hassani$^\star$ \thanks{$^\diamond$Fondazione
    Bruno Kessler, via Sommarive, 18 I-38123 Povo, Trento, Italy; $^\star$CERI/LIA, University of Avignon, 339, Chemin des Meinajaries,
Avignon, France; $\ddagger$ Nokia Bell Labs, Paris-Saclay, France. This research was
performed while the first author was visiting Nokia Bell Labs.}}
\maketitle
\begin{abstract}
High availability of containerized applications requires to perform robust storage
of applications' state. Since basic replication techniques are extremely costly at
scale, storage space requirements can be reduced by means of erasure and/or repairing codes.

In this paper we address storage regeneration using repair codes, a robust distributed 
storage technique with no need to fully restore the whole state in case of failure. In
fact, only the lost servers' content is replaced. To do so, new clean-slate storage
units are made operational at a cost for activating new storage servers and a cost for
the transfer of repair data. 
  
Our goal is to guarantee maximal availability of containers' state files by a given deadline. 
Upon a fault occurring at a subset of the storage servers, we aim at ensuring 
that they are repaired by a given deadline.  We introduce a controlled fluid model and derive 
the optimal activation policy to replace servers under such correlated faults. 
The solution concept is the optimal control of regeneration via the Pontryagin minimum principle.
We characterize feasibility conditions and we prove that the optimal policy is of threshold type.
Numerical results describe how to apply the model for system dimensioning and show the tradeoff
between activation of servers and communication cost. 
\end{abstract}

\section{Introduction}\label{sec:intro}

Container technology has quickly become the most promising cloud virtualization technique 
for it is lightweight and portable to different hardware. The uptake of containerization 
is fast up to the point that containers have become the unique runnable entities supported by 
Google's infrastructure \cite{BurnsComACM2016}. The main difference of containers with 
respect to traditional virtual machines is the fact they are executed in the 
application space of a server. In fact, container's deployment does not require the 
instantiation of a full operating system on top of the one ruling the host server, 
thus representing a lighter solution with faster setup time. 

However, performing {\em high availability} of containerized applications is still a developing concept,
e.g., building blocks such as failure detection and failover management are missing~\cite{LiIC2E2015}.
Virtual machines and containers, in turn, may be supported by availability guarantees~\cite{Salapura2015}
corresponding to specific service level agreements (SLA) to remain continuously functional (staying operational
$99.999\%$ of the time is called the five nines rule~\cite{Gray1991}). 

High availability requires a large degree of fault tolerance, both at the software and the hardware level. 
In the case of containerized applications, whenever a container fails, such failure 
can be masked, while the related traffic and tasks are redirected to healthy replicas.
Incidentally, this is also the standard technique for seamingless migration of
containerized applications across cloud servers for load-balancing purposes.

Cloud native applications to be containerized are ideally instantiated in a stateless fashion.
This makes it simple to render container execution highly available. However, 
containerized applications not always can be made fully stateless. Instead, they
can store the running state in a replicated distributed storage. One existing
deployment in the literature is found  in~\cite{infinity}. 
By using dedicated plug-ins, persistent volume from inside 
containers is made accessible. The state is hence saved onto the distributed
file system before replacement or migration, and the new container can finally
access the recorded state \cite{LiIC2E2015}. 

In order to maintain an up-to-date version for restoring or to migrate running 
containers, snapshot images of the containers' state have to be created. Commit 
commands available on container platforms \cite{Docker} can be used and several 
optimizations are possible to this respect, e.g., by continuously synchronizing 
changes only. 
Furthermore, in this context many core aspects are relevant, including load balancing, replica
synchronization, system monitoring, alarm generation, and configuration management.
Such aspects are beyond the scope of this work. Instead, we focus on the mechanisms
for failure recovery of storage serves.

In fact, robustness of data storage becomes the bottleneck to ensure high availability for containers' 
state maintenance. Data loss events in data centers are reported as a common event by several operators, e.g,
FaceBook \cite{Borthakur2011} and Yahoo \cite{Chansler2012}. The traditional solution is to perform server
content replication using three-way random replication, considered the standard good practice in distributed
filesystem management~\cite{Ghemawat2003,Lakshman2010,Cidon2013,Cidon2015}. 

In the literature on distributed storage, nevertheless, there exist techniques to reduce redundancy, e.g., by 
means of erasure codes or by repairing codes. Erasure codes can achieve great savings in storage space, and 
are actually used by major cloud provides such as Facebook \cite{Rashmi2013} and Google \cite{Ghemawat2003}. 

The basic idea with erasure codes is that a file is split into $k$ chunks, and then encoded into $n=k+h$ chunks.
In case of $r\leq h$ server failures, the system state can be recovered by transferring the chunks from $k$ of the
$n-r$ remaining servers and decoding those to retrieve the whole original file. Then, the file can be encoded all over again
into $n$ chunks and finally the lost encoded chunks are restored on a set of $r$ replacement servers. We observe 
that in our context the servers may be either physical servers or virtual storage units, and faults may be 
due to simultaneous node failures due, e.g., to cluster-wide power outages \cite{Cidon2013}.

When there exists a large number of containers, the data transfer phase can become bottleneck 
for fast recovery in private clouds and a costly service to offer at scale in a public cloud. A recent solution
is  the usage of {\em repairing codes}~\cite{ShahRKR12,Dimakis,Sathiamoorthy2013}. Several trade-offs for such
technique are addressed in~\cite{Jiekak2013}, showing a $10$-fold improvement is possible over standard erasure coding.  

In this work, we investigate feasibility and cost of regeneration operations using repair codes under correlated
faults, i.e., when several servers fail at once. State availability requirements are represented by a deadline $T$ to
regenerate all servers. The cost that it takes to maintain seamless operation of containers' involve  both state
storage, i.e., activating enough replacement servers, and communication costs, i.e., the  cost of {\em transferring}
coded data chunks to regenerate lost servers. In the rest of the paper, the limit performance of the system are derived
using an optimal control framework. 

The paper is organized as follows. In Sec.~\ref{sec:related} we review the related literature, whereas
in Sec.~\ref{sec:system} we introduce the system model. In Sec.~\ref{sec:problem} we formulate the problem 
of state storage regeneration in the framework of optimal control. Sec.~\ref{sec:solution} details the solution.
Sec.~\ref{sec:num} provides numerical results and Sec.~\ref{sec:concl} concludes the paper. The complete proofs
of the statements derived in this paper can be found in Appendix.

\begin{table}[!t]\caption{Main notation used throughout the paper}
\centering
\begin{scriptsize}
\begin{tabular}{|p{0.20\columnwidth}|p{0.68\columnwidth}|}
\hline
{\it Symbol} & {\it Meaning}\\
\hline
$B$ & state size \\
$\beta$ & repairing chunk size \\
$\C=(n,d,k)$ & repairing code \\
$c_1$&  cost per repair node activation \\
$c_2$& cost per transferred repair chunk bit\\
$u(t)$ & activation control \\
$\zeta$ & maximum activation rate\\
$\lambda$ & number of repair chunks transferred per second\\
$\mu$ & repair server failure rate  \\
$X_d(0)$ & number of operational repair servers at time $0$\\
$X_0(t)$ & number of newly activated repair servers at time $t$\\
$X_k(t)$ & number of repair nodes having $k$ servers at time $t$\\
\hline
\end{tabular}\label{tab:notation}
\end{scriptsize}
\end{table}

\section{Related Works}\label{sec:related}

Designing robust storage in the cloud is a classical problem. Random replication schemes appeared in the early Google filesystem~\cite{Ghemawat2003}
and in Facebook data centers~\cite{Lakshman2010}. 
Basic erasure codes achieve higher reliability compared to replication with same storage~\cite{Weatherspoon2002}. The cost reduction in datacenter
footprints operations is dramatic, exceeding $50\%$, thus recommending their usage in next generation systems \cite{Khan2012RethinkingEC}. 
Hence, new specialized erasure codes appeared, such as the local reconstruction codes in Windows Azure Storage.~\cite{Huang2012}, or piggybacked
Reed-Solomon codes to reduce cross-racks restoration bandwidth in Facebook's datacenters~\cite{Rashmi2013}. 
The breakthrough in the field are the erasure codes introduced by Papailiopoulos and Dimakis  in~\cite{Dimakis}, a class of locally repairable codes
of maximum distance type separable (MDS). Several follow up works, e.g., \cite{ShahRKR12,Sathiamoorthy2013} have explored the fundamental tradeoff of
such codes. They can be either of the minimum storage (MSR) or the minimum bandwidth (MBR) regenerating type. 
When the code can be  maintained in systematic form, simple {\em repair by transfer} with no decoding operations is possible. However, in general,
regeneration involves also decoding and so computing-time~\cite{Jiekak2013}, a facet of the problem that we leave as part of future works.  

In the rest of the paper, we consider an assigned deadline for failsafe operations, as proposed in 
\cite{Salapura2015}: in that work, recovery time limits are imposed on the parallel failover of virtual machines based on customers' SLA plans.
Also, in this work we adopt a system perspective close to~\cite{Jiekak2013}. To the best of the authors' knowledge, this is the first paper
describing optimal control of failsafe operations for storage regeneration.

\section{System Model}\label{sec:system}

\begin{figure}[!t]
\centering
\includegraphics[width=0.40\textwidth]{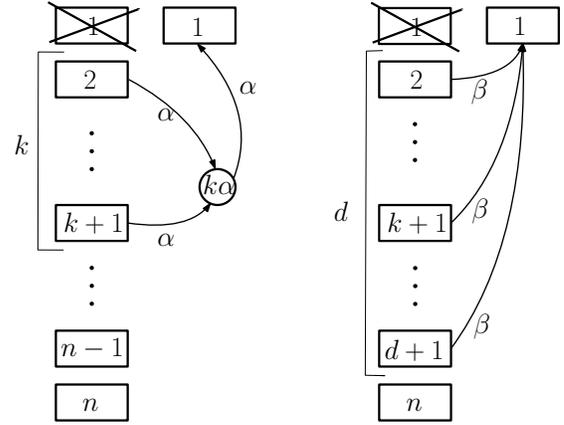}
\caption{Storage regeneration: repairing failures via erasure codes (left) and repairing codes (right); $r=1$, $d=n-2$}
\label{concept}
\end{figure}

In order to perform repair coding, the containers' state is divided into $k$ chunks and encoded into $n=k+h$ ones, by
using a repairing code $\C=(n,k,d)$, where $n> d > k$. Parameter $d$ represents the number of chunks that can be used
to repair a lost or corrupted one. Each chunk is hence stored by distributing the encoded chunks to $n$ servers. At
time $t=0$, $r$-servers fail, with $0 < r \leq n-d$, whereas $n-r$ servers are still operational. In case of a $r$-servers
fault, there are two main restoration options: either {\em full restoration} or {\em regeneration} of failed servers.
If $r<h$, full state restoration is possible from any set of $k$ servers chunks: full restoration requires to transfer
$k$ data chunks, which have $\alpha$ bytes each, to reconstruct the whole state file, to perform the encoding process all
over again and, finally, to transfer the re-encoded chunks to the destination servers (see Fig.~\ref{concept}). 

Instead, selective regeneration of failed servers is possible when $r<n-d$: each lost server is replaced by using the
chunks of $d$ repair servers, by transferring $\beta$ bits of information from each encoded chunk. Clearly, repairing
is possible as long as there exists at least $d$ {\em repair servers}. Optimal repairing MSR codes set $\alpha=B/k$
and $\beta=\alpha/(d-k+1)$, whereas optimal repairing MBR codes set $\alpha=2dB/[k(2d-k+1)]$ and
$\beta=\alpha/[k(2d-k+1)]$~\cite{ShahRKR12}.

In order to obey to availability constraints, we assume that repairing operations need to complete by time horizon
$T$, i.e., it must hold $X_d(T)= n$. Once the regeneration procedure through repair codes is completed, the full set of $n$
operational repairing nodes is restored. We model such procedure as follows. First, new repairing servers are activated,
e.g., by adding a new physical node to the datacenter, or by installing dedicated storage virtual machines on servers
already part of the fabric. They can be switched on at a maximum rate $\zeta$; the activation process is a Poisson process
with rate $\zeta$, i.e., new servers can be activated at rate $\zeta>0$ new replacement servers per second. 

Once activated, a repairing server downloads parity information from $d$ operational repairing servers. We assume that each
chunk transfer requires an exponential random time with mean $1/\lambda>0$. The regeneration procedure has two cost components:
\begin{itemize}
\item[i.]{\em activation cost}: activating a new repairing server has a cost $c_1$ per repairing server, due to the usage of
legacy hardware in the datacenter and the related setup costs;  
\item[ii.]{\em transfer cost}: data transfer has a cost $c_2$ per bit, hence a chunk transfer has a cost $c_2 \beta$.  
\end{itemize}
During the regeneration process, due to hardware and/or software issues, failure of repairing  servers may occur as well;
failure instants are modeled as exponential random variables of parameter $\mu$.

The number of newly activated servers is denoted by $X_0(t)$, whereas $X_k(t)$ denote the number of 
replacement servers that have $k$ repair chunks, for $k=1,\ldots,d$. Only nodes retrieving $d$ chunks
are operational replacement nodes: for notation's sake, we shall consider $X_d(t)$ the whole set
of repairing nodes, i.e., those include the $n-r$ which have not crashed. Restoration of the system 
using repair codes is possible if and only if $X_d(t) \geq d$ at each point in time (if $k \leq X_d(t)<n-d$ 
only full restoration is possible, if $X_d(t)<k$, containers' state is lost.).

\subsection{Markov model and fluid approximation}

We shall study how to optimally activate new repairing servers in order to successfully
restore all $n$ servers within finite time horizon $T$ at minimum cost. We start by assuming a stochastic
control, namely, the probability $u$ that a replacement server is activated. The activation rate of new repairing
servers is $\zeta \cdot u(t)$. The control acts by thinning the maximum activation rate $\zeta$, which can be easily implemented
by randomly sampling servers to be activated. Thus, $\zeta \cdot u(t)$ is the rate at which replacement servers become
active subject to stochastic control $u(t)$. Let us define the state of the system as $\bX=(X_0,X_1,\ldots,X_d)$,
where $X_k$ denotes the number of servers which have retrieved the content from $k$ repairing servers. The state $\bX(t)$
has a dynamics described by a continuous time Markov decision process (MDP), where we observe that all states $\bX$ such
that $X_d< d$ are absorbing, since no repairing is possible.

Let assume that once $k$ chunks are acquired, the repair process proceeds by downloading from the remaining $d-k$
repairing servers. Hence, for any initial state $\bx$, we can write the entries of the transition probability matrix 
\begin{eqnarray}\label{eq:mdp}
&&\hskip-12mm P_{\bx',\bx}(dt)=\Prob{\bX_{t+dt}=\bx'|\bX_t=\bx}=\nonumber\\
&&      =\begin{cases}
  				  \zeta \, u(t) \, dt & \mbox{if} \; \bx ' = \bx+\e0 \\
 			      \mu \, x_0 \, dt   & \mbox{if} \; \bx ' = \bx-\e0  \\
   				 (d-k+1)\lambda x_{k-1}\, dt  & \mbox{if}\; \bx ' = \bx+\e{k}-\e{k-1} \\
    			  \mu x_k \, dt & \mbox{if} \; \bx ' = \bx-\e{k-1}   \\
   				 o(dt) & \mbox{otherwise}  
                 \end{cases}
\end{eqnarray}
where with $\e{k}$ is the $k$-th element of the standard basis. The first row describes the
event of activation and the second row the failure of a newly activated repairing server,
respectively. The third row describes the acquisition of a repair chunk by a repairing node
having $k-1$ chunks, and the fourth row describes the failure of a node having retrieved $k$
chunks. The last row states that multiple transitions are negligible in the corresponding
infinitesimal generator.

The process of regeneration of the servers can be studied using a fluid model. Due
to the structure of system \eqref{eq:mdp}, the meanfield approximation can be proved
tight for $n$ in the order of a few tenths \cite{lucile}. By using the resulting fluid 
approximation, in the next section we shall obtain an optimal control problem in continuous time.

The control space $\U$ is the set of the piecewise continuous
functions taking values in $[0,1]$. The dynamics of the number of repairing servers thus writes 
\begin{eqnarray}\label{eq:dynamics}
&&\hskip-5mm\dot X_0(t)=  -\mu_0 \, X_0(t) + \zeta \, u(t) =f_0(X,u,t)  \nonumber  \\
&&\hskip-5mm\dot X_1(t) = -\mu_1 \, X_1(t) + d\lambda X_0(t) =f_1(X,u,t) \nonumber  \\
&&\qquad \vdots \nonumber  \\
&&\hskip-5mm\dot X_k(t) = -\mu_k \, X_k(t)+ (d-k+1)\lambda X_{k-1}(t) =f_k(X,u,t) \nonumber  \\
&&\qquad \vdots \nonumber  \\
&&\hskip-5mm\dot X_d(t)= -\mu_d \,  X_d(t)+ \lambda X_{d-1}(t) =f_d(X,u,t)
\end{eqnarray}
The ODE system \eqref{eq:dynamics} represents the dynamics of the regeneration
process. Here, $\mu_k=\mu+\lambda(d-k)$ is the rate at which servers with $k$ chunks
fail to repair plus the rate at which they receive a new chunk, thus joining those
having $k+1$ chunks. Also, the first equation of the ODE system \eqref{eq:dynamics},
namely $f_0(\cdot)$, incorporates the activation of new peers at controlled rate
$\zeta \, u(t)$.

\section{Optimal Control Problem}\label{sec:problem}

The objective is to minimize the cost to restore the system by deadline $T$: the storage regeneration
dynamics~\eqref{eq:dynamics} is controlled by activation control $u$. Hence, the objective function writes
\begin{equation}\label{eq:objective}
J(u)= \int_0^T \left [ c_1 \zeta u(v) + c_2 \, \beta \sum_{i=0}^{d-1} \lambda (d-i) \, X_i(v) \right ] dv
\end{equation}
where the first term appearing in the integral is the servers' activation cost whereas the second one is the 
cost for transferring chunks to repair servers.  We shall solve the following optimization problem:
\begin{prob}[Optimal Storage Regeneration]\label{prob:problem}
Find a control policy $u$ which solves:  
\begin{eqnarray}\label{eq:maximization}
&&\min_{u \in \mathcal U} \;J(u)\nonumber \\
  &&\mbox{s.t.} \quad  X_d(t)\geq d\qquad\;\forall \; 0 \leq t \leq T  \\
  &&            \hskip4.5mm \quad  X_d(T) = n  \nonumber 
\end{eqnarray}
where $d \leq X_d(0) \leq n$.
\end{prob}
In order for the repairing procedure to succeed, at least $d$ repair nodes must
be present at all points in time. We observe that, because \eqref{eq:dynamics}
describes the deterministic dynamics of the mean value of the underlying MDP,
it is possible that some sample paths do not satisfy the constraints, an event
that should occur with small probability.  To this aim, is possible to tighten
constraints appearing in \eqref{eq:maximization}, in the form
\begin{eqnarray}\label{eq:margins}
&&d' = (1 + \epsilon_1) d \qquad n' = (1 + \epsilon_2) n,   \nonumber 
\end{eqnarray}
where $\epsilon_1,\epsilon_2>0$ represent relative margins. In the rest of the paper, we shall
refer to the case $\epsilon_1=\epsilon_2=0$ without loss of generality.

Hereafter, we shall determine the conditions when the problem is feasible, i.e., the set of solutions of the
problem is not empty. Actually, we recall that, as long as $k$ chunks exist in the system, full restoration is
still possible. However, we focus solely on the cases when regeneration is feasible, which can be determined
easily by analysis of the uncontrolled dynamics, as discussed next.


\subsection{Feasibility and System Dimensioning}\label{sec:feasib}


Let us denote $\overline X_d(t)$ the dynamics corresponding to $u(t)\equiv 1$ in the interval $[0,T]$.
Because the activation control is basically slowing down the maximum activation rate $\zeta$,
it holds $X_d(t)\leq \overline X_d(t)$ for all $t\in[0,T]$. Hence, it is immediate to observe that the problem is
feasible if and only the dynamics of $\overline X_d$ is compatible with the constraints. Such condition can be
derived in closed form. By writing the Laplace transform of \eqref{eq:dynamics}, i.e., $X_k(s)=\mathcal L\{X_k(t)\}$ we obtain

\begin{small}
\begin{equation}
\overline X_0(s)=\frac\zeta{s+\mu_0}, \overline X_1(s)=\frac{\overline X_1(s)}{s+\mu_1}, \ldots, \overline X_d(s)=\frac{X_{d-1}(s)+\overline X_d(0)}{s+\mu_d} \nonumber
\end{equation}
\end{small}

\noindent which in turn provides $\overline X_d(s)=\frac{\lambda^d d! \zeta}{\prod_{k=0}^d (s+\mu_k)}+ \frac{\overline X_d(0)}{s+\mu}$. As showed in the Appendix,
the following closed form expression for the dynamics of the  repairing servers holds:
\[
\overline X_d(t)=e^{-\mu t}\Big ( \zeta \big (1-e^{-\lambda t}\big )^d+\overline X_d(0)\Big )
\]
Feasibility conditions can be described in terms of the system parameters as follows:
\begin{lemma}\label{lem:feasibility}
Problem~\ref{prob:problem} is feasible if and only if $\zeta \big (1-e^{-\lambda T}\big )^d \geq n\,e^{\mu T} - \overline X_d(0)$ and it is
so for any $\mu\leq \overline \mu$, where
\[
\overline \mu := \min\{\overline \mu_n,\overline \mu_d\} 
\]
and $\overline \mu_n=\max\{\mu \geq 0 | {\overline X}_d(T)\geq n\}$ and  $\overline \mu_d=\max\{\mu \geq 0 | \min_{t\in[0,T]} {\overline X}_d(t) \geq d\}$.
\end{lemma}

In the rest of the paper we assume $\mu>0$ and feasibility in the sense meant by the previous statement. 

{\em System dimensioning.} Lemma~\ref{lem:feasibility} provides indications for dimensioning the system
in order to guarantee feasible regeneration. In particular, in the worst case we would need to transfer $n-d$ chunks
to newly activated repair nodes. In turn, one would choose the time horizon by which to repair, namely $T$,
and $\lambda$, i.e., the rate at which chunks can be transferred, and the code's triple $\C=(n,k,d)$, such in a way to
satisfy the assumptions of the above statement. 


\subsection{Relaxed problem}\label{sec:subaugment}


{\em Constraint Relaxation.} The terminal state constraint can be accounted by relaxing the problem
in the form 
\begin{eqnarray}\label{eq:objectivemod}
J_\gamma(u)= J(u) + \gamma \, (n - X_d(T))
\end{eqnarray}
by means of the terminal cost function $q(\bX):= \gamma \, (n - X_d(T))$. We note that
$\gamma \geq 0$ has the role of a multiplier, and when the constraint is active $\gamma>0$.

{\em State Augmentation.} In order to account for the first constraint, we operate the augmentation of the state space by
introducing an auxiliary variable
\[
\dot X_{d+1}(t) = (X_d(t)-d)^2\1{d - X_d(t)}
\]
where the indicating function $\1{x}=1$ if $x>0$ and $\1{x}=0$ if $x<0$. Since
\[
 X_{d+1}(t) = \int_0^T  X_{d+1}(v)dv + X_{d+1}(0).
\]
We impose the auxiliary constraint $X_{d+1}(T)=X_{d+1}(0)=0$: because $X_{d+1}(t) \geq 0$ for $t \in [0,T]$,
when such two constraints are satisfied, then $X_d(t) \geq d$ all over the interval $[0,T]$.

We denote the problem of minimizing $J_\gamma(u)$ the {\em relaxed problem} and it will be solved next.


\subsection{\em Hamiltonian formulation and Pontryagin Principle.}


Let denote $g(X,u,t)$ the instantaneous cost appearing inside the integral cost \eqref{eq:objective}.
In order to solve the optimal control problem, it is possible to write the Hamiltonian for the optimal control problem
in standard form
\[
 H(\bX,u,\bp)= p(t) \, f(\bX,u)+g(\bX)\nonumber
\]
where $p$ is the vector of co-state variables  Hence, according  to the Pontryagin Minimum Principle \cite{leitmann,Kirk},
the optimal control $u$ needs to satisfy
\begin{equation}
  u(t)=\arg\min_{\hskip-5mm u\in \mathcal U} H(X,u,p)\nonumber
\end{equation}
where the associated Hamiltonian system is  
\begin{eqnarray}\label{eq:system}
  \dot X_k &&= H_{p_k}(\bX,u,\bp) \\
  \dot p_k &&= - H_{X_k}(\bX,u,\bp)
\end{eqnarray}
We have $d+1$ terminal conditions in the form  $p_k(T)=q_{_{X_k}}(T)=0$ for $k=0,1,\ldots,d-1,d+1$. Also, terminal condition
$p_d(T)=q_{_{X_d}}(T)=-\gamma$ holds. 

\section{Solution}\label{sec:solution}

In order to solve the storage regeneration problem, we can write the Hamiltonian as 
\begin{eqnarray}\label{eq:hamiltonian}
  && \hskip-6mm H(\bX,u,\bp)=\zeta \big ( c_1 +  p_0(t)\big ) \, u(t) -\mu X_0(t)p_0(t)+ \nonumber \\
    && + c_2 \beta \sum_{i=0}^{d-1} \lambda (d-i) \, X_i(t) \nonumber \\ 
    && +\sum_{k=1}^d \Big [  -\mu X_k(t) +\lambda (d - k + 1) \cdot X_{k-1}(t) \Big ] p_k(t) \nonumber \\
    && +  p_{d+1}(t)\,  (X_d(t)-d)^2\, \cdot \1{d - X_d(t)}
\end{eqnarray}
We can hence derive from \eqref{eq:system} the adjoint ODE system in the costate variables
\begin{eqnarray}\label{eq:adjoint}
&&\hskip-4mm\dot p_0 =-H_{X_0}= \mu_0    \cdot p_0 - \lambda d \cdot p_1 - c_2 \beta \lambda d \\
&&\hskip-4mm\dot p_1 =-H_{X_1}= \mu_1    \cdot p_1  - (d-1) \lambda \cdot p_2 - c_2 \beta \lambda (d-1)\nonumber\\
  && \qquad \vdots \nonumber\\
&&\hskip-4mm\dot p_k =-H_{X_k}= \mu_k    \cdot p_{k}  - (d-k) \lambda \cdot p_{k+1} - c_2 \beta \lambda (d-k)\nonumber\\  
  && \qquad \vdots \nonumber\\
&&\hskip-4mm\dot p_{d-1} =-H_{X_{d-1}}= \mu_{d-1}  \cdot p_{d-1}  - \lambda \cdot p_{d} - c_2 \beta \lambda \nonumber\\
  &&\hskip-4mm\dot p_{d} =-H_{X_d}= \mu_d p_{d} 
  - 2(X_d(t)-d) \cdot \1{d - X_d(t)} p_{d+1}\nonumber\\
&&\hskip-4mm\dot p_{d+1}=0 \nonumber
\end{eqnarray}
In what follows, we will derive the structure of the solutions of the optimal control problem. A 
{\em bang-bang policy} \cite{leitmann,Kirk} is one where $u(t)$ takes only extreme values,
that is $u(t)=1$ or $u (t)=0$ a.e. in $[0,T]$. 

Notice that bang-bang policies are very convenient for implementation  purposes since they rely
only on a set of {\em switching epochs}, where the  control  switches from $1$ to $0$ or {\em vice versa}.
A {\em threshold policy} is one in the form 
\begin{equation}\label{eq:thresh}
  u(t)=\begin{cases}
   0 & \tson < t \leq T \cr
  1 & 0 < t \leq \tsoff \cr
  0 & \tsoff < t < T \cr
  \end{cases}
\end{equation}
Threshold policies are convenient since they depend on a pair of parameters only, namely thresholds  $\tson$ 
and $\tsoff$. 

{\em Bang-bang structure.} We observe that \eqref{eq:hamiltonian} is linear in the control $u$.
Hence, because the optimal activation control minimizes the Hamiltonian, the optimal policy has to satisfy 
\begin{equation}\label{eq:switching}
u(t) =
\left\{
\begin{array}{lr}
1 & \mbox{ if }   p_0(t) < -c_1 \\
0 &  \mbox{ if }  p_0(t) > -c_1
\end{array}
\right.
\end{equation}
which depends on the dynamics of $p_0$, i.e., of the ODE system \eqref{eq:adjoint}. Actually,
in order to prove that the policy is bang-bang and non-degenerate, we need also
to prove that the policy has a finite number of switches and that there are no singular arcs,
i.e., no arcs where the Hamiltonian is null over an interval of positive measure. 

\begin{lemma}\label{thm:singulararcs}
If the problem is feasible, the optimal policy is bang-bang with no singular arcs.
\end{lemma}
The dynamics of $p_0$ can be derived in closed form:
\begin{lemma}\label{thm:multiplier}
  It holds $p_0(t)=-F(t)+G(t)$ where
  \begin{eqnarray}
    &&\hskip-2mm\displaystyle F(t)=\gamma \Big (1 - e^{-\lambda (T-t)}\Big )^d e^{-\mu (T-t)} \nonumber \\
    &&\hskip-2mm\displaystyle G(t)=c_2 \beta d \lambda \sum_{k=0}^{d-1} \binom{d-1}{k} \int_0^{T-t} (e^{\lambda v} -1)^k e^{-(\mu+\lambda d)v} d v \nonumber
  \end{eqnarray}
\end{lemma}
Next, we characterize solutions of the relaxed problem which correspond to feasible solutions.


\subsection{Pure Activation Cost}\label{sec:pureact}


We start our analysis from the simpler case when the transfer cost is negligible compared to the activation cost, i.e.,
$c_2=0$. It is hence possible to derive explicit relations on the structure of the optimal control.
\begin{thm}\label{thm:thm1}
If $c_2=0$, then a solution of the relaxed problem is a threshold policy, in particular:\\
\noindent i. Single switch: $\tson=0$ and $0< \tsoff < T$ iff $ \mu \leq \mu_0$; \\
\noindent ii. Null control: $0=\tson=\tsoff$ iff $\mu > \mu_0$, and $m \geq -c_1$, where $m=\min_{v\in[0,T]}\{p_0(v)\}$;\\
\noindent iii. Double switch: $0<\tson<\tsoff\leq T$ iff $\mu > \mu_0$, and $m <-c_1$ \\
The critical value
\[
\mu_0:= \max \{0,\frac dT \log (\sqrt[d]{{\gamma}/{c_1}}\, (1-e^{-\lambda T})) \}
\]
while the switching epochs write $\tson=\max\{0, T+\frac 1{\lambda}\log\zon\}$, $\tsoff=T+\frac 1{\lambda}\log\zoff$,
where $\zon \leq \zoff$ are the two solutions for $0\leq z \leq 1$ of the equation
\[
(1-z) = \sqrt[d]{\frac {c_1}{\gamma}}z^{-\frac{\mu}{\lambda d}}
\]

\end{thm}


\subsection{General case}\label{sec:transfer}


In the general case, it is sufficient to characterize the dynamics of the multiplier $p_0(t)$ in terms of the extremal points attained
in the interior of $[0,T]$.
\begin{lemma}\label{lem:extrema}
Let $\S(\gamma)$ be the  set of the interior extremal points of $p_0(t)$ for a given choice of the constraint multiplier
$\gamma$. Then, $\S(\gamma)$ is one of the following forms: $\emptyset$, $\{M\}$, or $\{m,M\}$, where $m:=p_0(t_m)$
denotes a minimum and $M:=p_0(t_M)$ a maximum, and it holds $0\leq t_m<t_M < T$.
\end{lemma}
Finally, as proved in the Appendix.
\begin{thm}\label{thm:thm_str}
The optimal solution of the relaxed problem is a threshold control.
\end{thm}

The optimal control is hence a threshold policy for which the presence of an initial delay, i.e., $\tson>0$, 
depends on the parameters of the system. However, as a straightforward application of the optimality principle,
given an optimal threshold policy with $\tson$ and $\tsoff$, for a given pair $T$ and $r$, the new threshold policy
where $\tson'=0$, $\tsoff'=\tsoff-\tson$ is optimal for the problem where $r'=n-X_d(t_m)\geq r$ and horizon
$T'=T-\tson<T$. Thus we obtain the optimal solution in threshold form with no initial delay for more conservative
conditions, i.e., for smaller time horizon and larger number of failed servers, and yet having same cost. 

Note that, in the relaxed problem, we cannot exclude the null control $u \equiv 0$, i.e., when
$p(0)>-c_1$ and $m>-c_1$. But, it cannot solve the constrained problem: to do so
we need to determine the optimal multiplier $\gamma$, as seen next.


\subsection{Optimal multiplier}\label{sec:multiplier}


The discussion so far has addressed the relaxed problem, and the multiplier $\gamma$ has been treated
as a constant for the sake of discussion. However, determining the optimal solution requires to
identify a pair $(u^*,\gamma^*)$ where $u^*$ solves the original constrained problem. 

The main result in this section is that we can calculate the value $\gamma^*$ using
a simple bisection search as described in Algorithm~\ref{alg:algorit1}, under the 
feasibility assumptions of Lemma~\ref{lem:feasibility}. The algorithm
starts by exploring the interval for $\gamma\in [0,\gamma_0]$, where $\gamma_0>0$
is a suitably large value such that it holds $X_d(T)\geq n$. At line 5, 6 and 7
it solves the optimal control problem determining finally the terminal value $X_d(T)$
 within a certain tolerance $\varepsilon>0$. 

The search algorithm leverages the fact that the terminal number of repair servers
is monotone in $\gamma$. In fact, when the target value number exceeds $n$, it
explores on the left of the current interval, i.e., it searches in $[\gamma_L,\gamma\,]$.
Viceversa, when the target value is below $n$, it explores the right interval
$[\,\gamma,\gamma_R]$.

The formal justification of the correctness of the above search strategy, and the
optimality of the output of the algorithm is resumed by the following result, proved in the Appendix. 
\begin{thm}\label{thm:mult}
Under the assumptions of Lemma~\ref{lem:feasibility}, the optimal pair $(u^*,\gamma^*)$ which
solves the relaxed problem is unique, $u^*$ solves Prob.\ref{prob:problem}, and $\gamma^*$ 
can be approximated using a bisection search as in Alg.~\ref{alg:algorit1}.
\end{thm}


\begin{algorithm}[t]\caption{Optimal Regeneration Control}
\begin{small}
\begin{algorithmic}[1]
\small
\STATE\textbf{input:} {$T$, $\beta$, $\lambda$, $c_1$, $c_2$, $\varepsilon$}
\STATE\textbf{\hskip8mm} {$\gamma_0$ s.t. $u$ from \eqref{eq:switching} is such that $X_d(T)\geq n$}
\STATE\textbf{initialize:} {$\gamma_R \leftarrow \gamma_0$, $\gamma_L \leftarrow 0$, $i\leftarrow 0$}
\WHILE {$|X_d(T) - n|>\varepsilon$}
\STATE Step $i\leftarrow i+1$
\STATE $\gamma_i\leftarrow (\gamma_L+\gamma_R)/2$
\STATE Obtain $p_0(t)$, $t\in [0,T]$ solving backwards \eqref{eq:adjoint}
\STATE Calculate the optimal control $u_i$ according to \eqref{eq:switching}
\STATE Obtain $X_d(t)$, $t\in [0,T]$ solving forward \eqref{eq:dynamics}
\IF {$X_d(T)>n$}
\STATE
$\gamma_R\leftarrow \gamma_i$ 
\ELSE
\STATE $\gamma_L\leftarrow \gamma_i$
\ENDIF
\ENDWHILE
\RETURN $(u_i,\gamma_i)$
\label{algo1}
\end{algorithmic}\label{alg:algorit1}
\end{small}
\end{algorithm}

\section{Numerical Results}\label{sec:num}

\fboxrule=1pt
\fboxrule=1pt
\begin{figure*}[t]
\begin{minipage}{6cm}{\includegraphics[width=\linewidth]{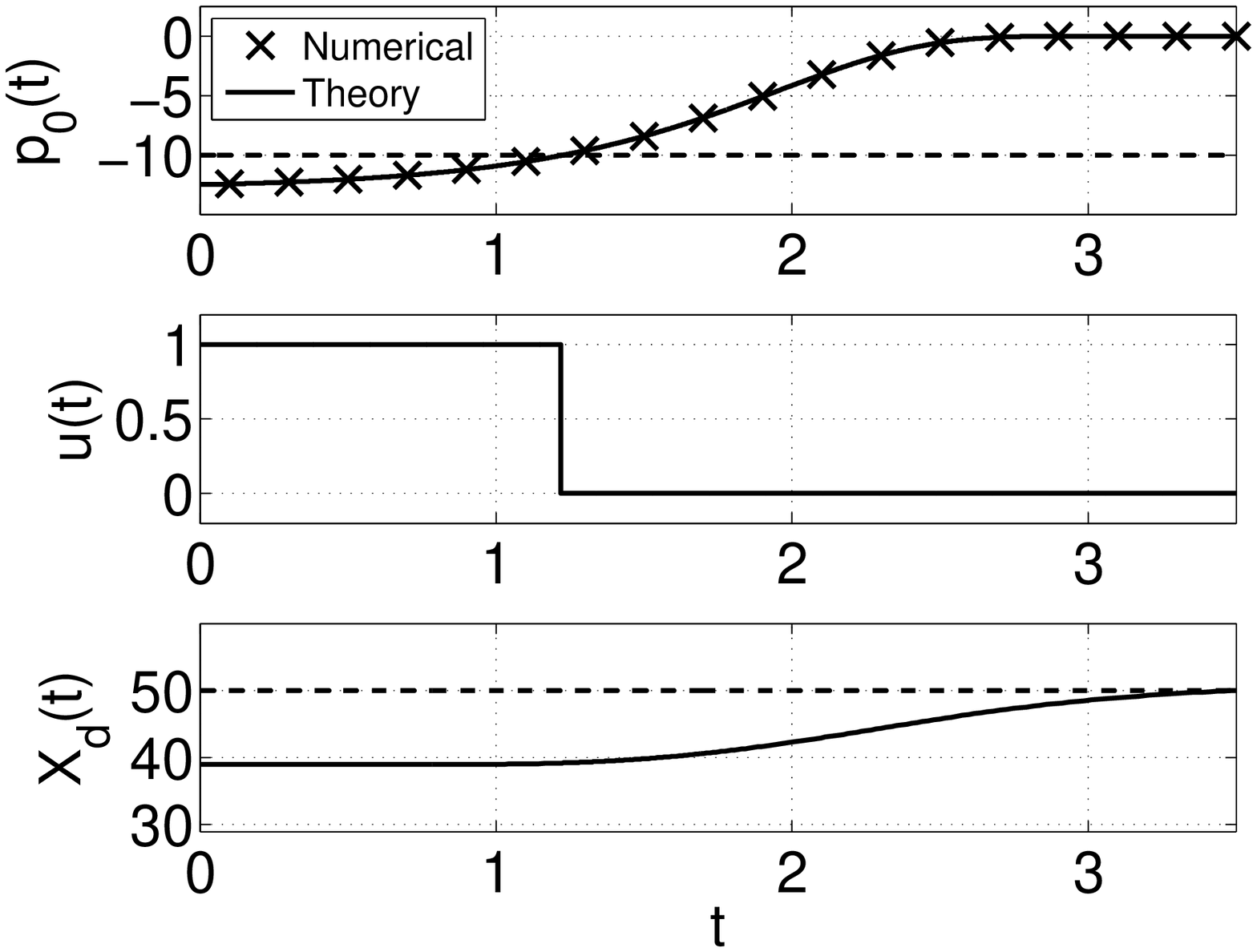}}\put(-170,130){a)}\end{minipage}
\begin{minipage}{6cm}{\includegraphics[width=\linewidth]{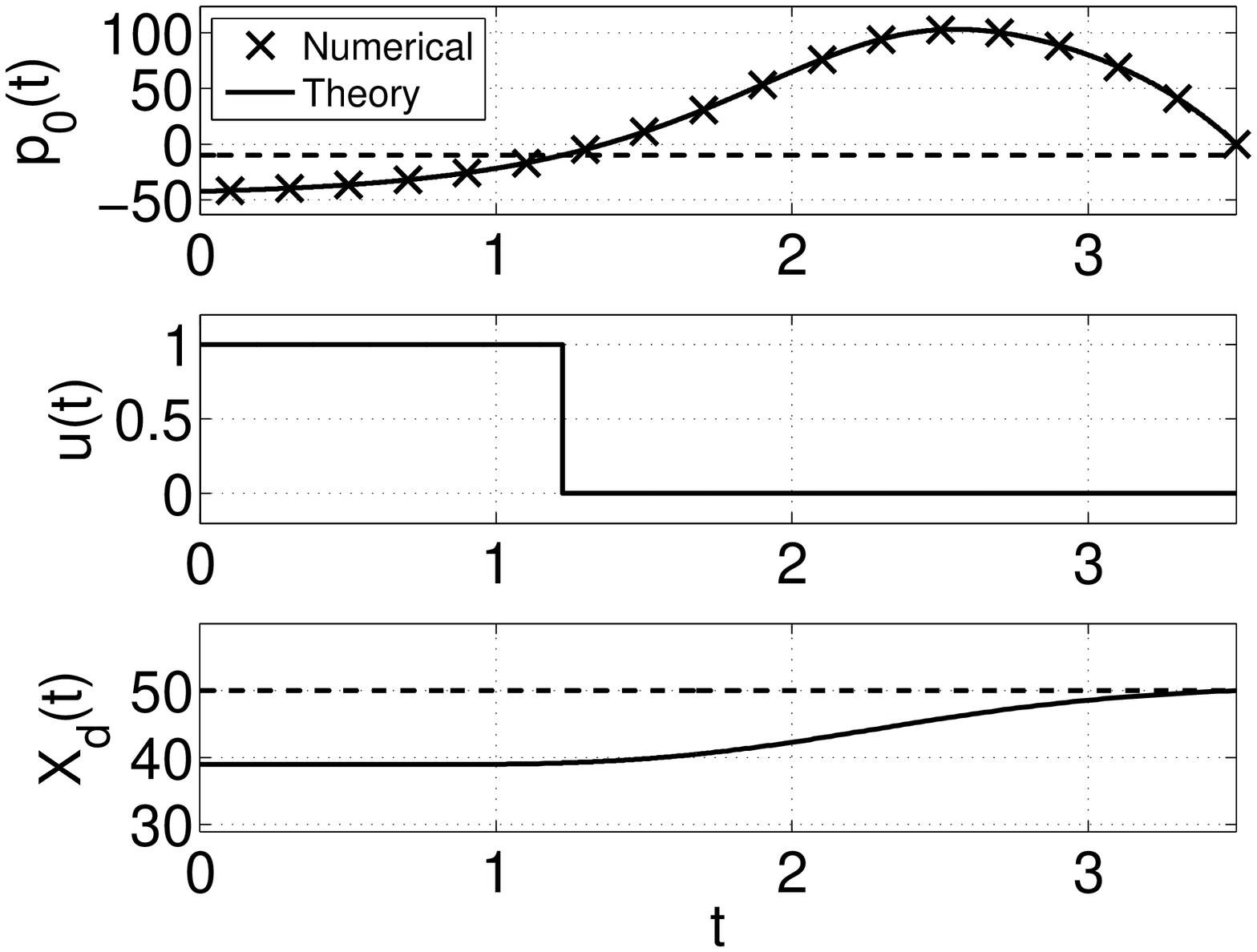}}\put(-170,130){b)}\end{minipage}
\begin{minipage}{5cm}
\begin{tabular}{c c || c  c  c }
   \multicolumn{2}{c}{\multirow{2}{*}{\fcolorbox{gray}{lightgray}{$J^*(u^*)$}} }          & \multicolumn{3}{c}{$c_2$}          \\ 
    &          &   \em \small 0    &   \em \small 10     &           \em \small 100          \\\hline\hline
   
  \multirow{3}{*}{$c_1$}& \em \small 1      & 12.2  &   169.0 &   1580.6  \\
        & \em \small 10      & 122.5  &  279.1 &   1691.9   \\
        & \em \small 20      & 244.9  &  401.3 &   1812.9   \\
\end{tabular}
\begin{small}
\begin{tabular}{c c || c  c  c }
   \multicolumn{2}{c}{\multirow{2}{*}{\fcolorbox{gray}{lightgray}{$\gamma^*$}} }          & \multicolumn{3}{c}{$c_2$}          \\ 
   &          &   \em \small 0    &   \em \small 10     &           \em \small 100          \\\hline\hline
  \multirow{3}{*}{$c_1$}&  \em \small 1  & 1.2766   & 17.5851  & 164.0627  \\ 
        & \em \small 10    &   12.8000  &  29.1024  & 175.8790  \\ 
        & \em \small 20   & 25.5990  &  41.7977  &  188.2813    \\ 
\end{tabular}
\end{small}
\end{minipage}
\caption{Optimal regeneration  control a) zero communication cost b) $c_2=100$ dollar/Gbyte  c) The optimal cost and the
optimal multiplier as function of costs $c_1$ and $c_2$.}\label{fig1}
\end{figure*}
This section presents some numerical results on optimal storage regeneration under a realistic parameter setting. It also
serves the purpose of explaining how to make use of the proposed model to characterize limit performance of the regeneration
technique under prescribed deadline constraints. We have assumed a reference $\C=(n,k,d)$ MBR repairing code. The parameters
of the code are $n=50$, $k=10$ and $d=20$ ~\cite{Jiekak2013}\footnote{In~\cite{Jiekak2013} the code redundancy targets
storage availability of $0.99$}.
Also, the reference container state size is assumed $B=10$ Gbytes. We recall that, based on the fundamental relation on MBR
codes, we can derive the chunk size as $\beta=2B/(k(2d-k+1))$~\cite{ShahRKR12}, which in this case amounts to
$\beta=64.5161$ Mbytes.

The numerical setting is completed by assuming that repairing servers may fail according to rate $\mu=0.001 s^{-1}$ (we
remind that in our model server failures during restoration are exponential random variables of parameter $\mu$).
Furthermore, the maximum rate at which repairing servers can be activated is set as $\zeta=10$ servers/s. Also, we
need to make assumptions on the available network throughput: in our scenario, the throughput available for repairing
operations is $1$ Gbit/s. This value matches link speeds of production datacenters: peak bitrates for repair chunks
transfer can be attained when performing restoration in priority, i.e., giving highest priority to the traffic
operating the transmission of repairing chunks. The resulting target horizon for repairing has been set to $T=3.5$ s,
which is feasible given the setting considered. 

Fig.~\ref{fig1}a and Fig.~\ref{fig1}b depict the results of the optimal activation control in case of simultaneous
failure of $r=11$ servers at time $t=0$. We have reported on the dynamics of the costate variable $p_0(t)$, superimposed to
the switching threshold value, namely $-c_1$ (upper graph), the graph of the corresponding optimal control dynamics (middle
graph) and the one corresponding to the dynamics of the number of repairing servers $X_d(t)$ (bottom graph).

In both cases, the optimal multiplier $\gamma^*$ has been determined using Algorithm~\ref{alg:algorit1} with tolerance
$\epsilon=0.05$. In particular, in Fig.~\ref{fig1}a we have considered the case of a null communication cost $c_2=0$,
which corresponds to $\gamma^*=12.7719$ whereas in  Fig.~\ref{fig1}b we have considered $c_2=100$ dollars/Gbyte, for
which the optimal cost is attained for $\gamma^*=175.855$. In both cases the threshold policy is such
that the pair $\tson=0$ s and $\tsoff= 1.22$ s identifies the unique control driving he dynamics to satisfy 
terminal state constraint $X_d(T)= n$.

Fig.~\ref{fig1}c contains two tables calculated for different values of the cost $c_1$ and $c_2$. They report on the value
of the optimal cost $J^*(u^*)$. We note that, as expected, it increases with both cost $c_1$ and $c_2$.
Also, we observe same behavior for $\gamma^*$: the optimal multiplier value increases and we ascribe this behavior
to the fact that the value of $\gamma$ has to enforce the terminal state constraint against augmented running costs $c_1$ and $c_2$.

\section{Conclusions}\label{sec:concl}

In this paper we have presented an analytical framework for the
optimal control of state regeneration, a promising technology in
order to offer high availability of containerized applications
at scale and ease stateful containers' migration. The
idea is that leveraging the network filesystem, it is possible
to decouple the storage of containers' state and the execution of
application images running in pods. 

We have studied optimal time-constrained regeneration, a crucial
aspect to ensure high availability in the containers' state access. 
Under failure of a number of servers, regeneration is performed
by transferring repairing chunks to newly deployed, clean
slate repair servers. This occurs at a communication cost and 
at a server activation cost. The optimal activation strategy is of 
threshold-type and can be evaluated in closed form.

This work has been motivated by the limited number of studies
on storage regeneration at system level~\cite{Jiekak2013} and it 
is by no means conclusive. Indeed, several research directions are due
in order to understand the potential of these novel
restoration techniques in cloud systems.

The first one relates to the frequency of updates of the containers'
state, a design choice required in order to decide how often to dump
the containers' state onto the network filesystem. Such rate
determines how much of the computation already elapsed can be
recovered using regeneration.

Another relevant issue is the case of repeated failures. Actually,
the information on where faults are more likely becomes available
to the administrator over time, e.g., based on direct observation
or online learning techniques. The optimal policy may in turn
span several cycles of faults/restorations and would account
for techniques to learn the aposteriori distribution of faults
over which to operate the optimal control.

Also, correlated faults described in this work are simultaneous. In reality,
they may be scattered in time, e.g., due to cascading failures.
Under such fault dynamics, the optimal control studied in this work
may be suboptimal. New models should identify how to counter the effect
of later additional faults occurring during regeneration.


\begin{thebibliography}{10}
\providecommand{\url}[1]{#1}
\csname url@samestyle\endcsname
\providecommand{\newblock}{\relax}
\providecommand{\bibinfo}[2]{#2}
\providecommand{\BIBentrySTDinterwordspacing}{\spaceskip=0pt\relax}
\providecommand{\BIBentryALTinterwordstretchfactor}{4}
\providecommand{\BIBentryALTinterwordspacing}{\spaceskip=\fontdimen2\font plus
\BIBentryALTinterwordstretchfactor\fontdimen3\font minus
  \fontdimen4\font\relax}
\providecommand{\BIBforeignlanguage}[2]{{%
\expandafter\ifx\csname l@#1\endcsname\relax
\typeout{** WARNING: IEEEtran.bst: No hyphenation pattern has been}%
\typeout{** loaded for the language `#1'. Using the pattern for}%
\typeout{** the default language instead.}%
\else
\language=\csname l@#1\endcsname
\fi
#2}}
\providecommand{\BIBdecl}{\relax}
\BIBdecl

\bibitem{BurnsComACM2016}
B.~Burns, B.~Grant, D.~Oppenheimer, E.~Brewer, and J.~Wilkes, ``{Borg, Omega,
  and Kubernetes},'' \emph{Comm. of the ACM}, vol.~59, no.~5, pp. 1837--1852,
  May 2016.

\bibitem{LiIC2E2015}
W.~Li and A.~Kanso, ``Comparing containers versus virtual machines for
  achieving high availability,'' in \emph{Proc. of IEEE IC2E}, Tempe, US, March
  9-12 2015.

\bibitem{Salapura2015}
V.~Salapura, R.~Harper, and M.~Viswanathan, ``{ResilientVM}: High performance
  virtual machine recovery in the cloud,'' in \emph{Proc. of ACM AIMC},
  Bordeaux, France, Apr 21-24 2015, pp. 7--12.

\bibitem{Gray1991}
J.~Gray and D.~P. Siewiorek, ``High-availability computer systems,''
  \emph{Computer}, vol.~24, no.~9, p. 39–48, 1991.

\bibitem{infinity}
{Infinit International Inc}, https://infinit.sh/documentation/reference.

\bibitem{Docker}
Docker, ``Docker: The linux container engine,'' http://www.docker.io.

\bibitem{Borthakur2011}
D.~Borthakur \emph{et~al.}, ``{Apache Hadoop} goes realtime at {Facebook},'' in
  \emph{Proc. of ACM SIGMOD PODS}, Athens, Greece, June 12-16 2011.

\bibitem{Chansler2012}
R.~J. Chansler, ``Data availability and durability with the hadoop distributed
  file system,'' \emph{The USENIX Magazine}, vol.~37, no.~1, February 2012.

\bibitem{Ghemawat2003}
S.~Ghemawat, H.~Gobioff, and S.-T. Leung, ``The {Google} file system,''
  \emph{SIGOPS Oper. Syst. Rev.}, vol.~37, no.~5, pp. 29--43, Oct. 2003.

\bibitem{Lakshman2010}
A.~Lakshman and P.~Malik, ``Cassandra: A decentralized structured storage
  system,'' \emph{SIGOPS Oper. Syst. Rev.}, vol.~44, no.~2, pp. 35--40, Apr.
  2010.

\bibitem{Cidon2013}
A.~Cidon, S.~Rumble, R.~Stutsman, S.~Katti, J.~Ousterhout, and M.~Rosenblum,
  ``Copysets: Reducing the frequency of data loss in cloud storage,'' in
  \emph{Proc. of USENIX ATC}, San Jose, US, June 26-28 2013.

\bibitem{Cidon2015}
A.~Cidon, R.~Escriva, S.~Katti, M.~Rosenblum, and E.~G. Sirer, ``Tiered
  replication: A cost-effective alternative to full cluster geo-replication,''
  in \emph{Proc. of USENIX ATC}, Santa Clara, CA, July 8-10 2015.

\bibitem{Rashmi2013}
K.~V. Rashmi, N.~B. Shah, D.~Gu \emph{et~al.}, ``A solution to the network
  challenges of data recovery in erasure-coded distributed storage systems: A
  study on the {Facebook} warehouse cluster,'' in \emph{Proc. of USENIX
  HotStorage}, San Jose, CA, June 27-28 2013.

\bibitem{ShahRKR12}
N.~B. Shah, K.~V. Rashmi, P.~V. Kumar, and K.~Ramchandran, ``Distributed
  storage codes with repair-by-transfer and nonachievability of interior points
  on the storage-bandwidth tradeoff,'' \emph{{IEEE} Trans. Information Theory},
  vol.~58, no.~3, pp. 1837--1852, 2012.

\bibitem{Dimakis}
D.~S. Papailiopoulos and A.~G. Dimakis, ``Locally repairable codes,''
  \emph{{IEEE} Trans. Information Theory}, vol.~60, no.~10, pp. 5843--5855, Oct
  2014.

\bibitem{Sathiamoorthy2013}
M.~Sathiamoorthy, M.~Asteris, D.~Papailiopoulos, A.~G. Dimakis \emph{et~al.},
  ``Xoring elephants: novel erasure codes for big data,'' in \emph{Proc. of
  PVLDB}, Riva del Garda, Italy, August 26-30 2013.

\bibitem{Jiekak2013}
S.~Jiekak, A.-M. Kermarrec, N.~Le~Scouarnec, G.~Straub, and A.~Van~Kempen,
  ``Regenerating codes: A system perspective,'' \emph{SIGOPS Oper. Syst. Rev.},
  vol.~47, no.~2, pp. 23--32, Jul. 2013.

\bibitem{Weatherspoon2002}
H.~Weatherspoon and J.~Kubiatowicz, ``Erasure coding vs. replication: A
  quantitative comparison,'' in \emph{Proc. of IPTPS}, Cambridge, MA, USA,
  March 7-8 2002.

\bibitem{Khan2012RethinkingEC}
O.~Khan, R.~C. Burns, J.~S. Plank, W.~Pierce, and C.~Huang, ``Rethinking
  erasure codes for cloud file systems: minimizing {I/O} for recovery and
  degraded reads,'' in \emph{Proc. of USENIX FAST}, San Jose, US, February
  14-17 2012.

\bibitem{Huang2012}
C.~Huang, H.~Simitci, Y.~Xu \emph{et~al.}, ``Erasure coding in {Windows Azure}
  storage,'' in \emph{Proc. of USENIX ATC}, Boston, MA, June 26-28 2012.

\bibitem{lucile}
E.~Altman, L.~Sassatelli, and F.~{De Pellegrini}, ``Dynamic control of coding
  for progressive packet arrivals in {DTN}s,'' \emph{IEEE Trans. on Wireless
  Comm.}, vol.~12, no.~2, pp. 725--735, 2013.

\bibitem{leitmann}
G.~Leitmann, \emph{An introduction to optimal control}.\hskip 1em plus 0.5em
  minus 0.4em\relax McGraw-Hill, 1966.

\bibitem{Kirk}
D.~{E. Kirk}, \emph{Optimal Control Theory. An Introduction.}, 13th~ed.\hskip
  1em plus 0.5em minus 0.4em\relax Prentice Hall, 2004.

\end{thebibliography}



\begin{appendix}

\section*{Proof of Lemma.~\ref{lem:feasibility}}
\begin{IEEEproof}
Feasibility is indeed equivalent to $\overline X_d$ to respect the constraints. Condition
$\zeta \big (1-e^{-\lambda T}\big )^d \geq d - \overline X_d(0)$ ensures that $\sup_\mu \overline X_d(T) \geq n$, which is
attained for $\mu=0$. We observe that $\overline \mu_d$ is also well defined: $g(\mu)=\inf_{t \in [0,T]}{\overline X}_d(t)$ is
a continuous function of $\mu$. Because $\inf_\mu g(\mu)=0$ and $g(0)=X_d(0)\geq d$, there exists a value of $\mu$ that satisfies
the definition. The statement follows immediately from the definition of $\overline \mu_d$  and $\overline \mu_n$ and from a continuity
argument.
\end{IEEEproof}

\section*{Proof of Lemma.~\ref{thm:singulararcs}}
\begin{IEEEproof}
Preliminarily, let observe that a feasible solution must be such that $X_d(t)\geq d$, for $t\in[0,T]$. Thus, 
the dual ODE system has to be solved as in the non-augmented case, where it holds $\dot p_{d} = \mu_d p_{d}$. Hence,
since the Hamiltonian is linear in the control, a feasible policy is a bang-bang one. In order to exclude the
presence of singular arcs, we need to exclude the possibility that $c_1 + p_0(t)=0$ over an interval $I$ of
positive measure. We shall prove that multiplier $p_0$'s cannot be a constant over any interval
$I$ of positive measure $\mu(I)>0$, and this guarantees that the control is actually bang-bang \cite{leitmann}. 
Let assume that $p_0$ is a constant $p_0 = -c_1$ over interval $I$: hence all its $k$-th order 
derivatives vanish in $I$. But, it follows from \eqref{eq:adjoint} that $p_1= \frac{\mu_0}{d \lambda}\, p_0$:
thus $p_1$ is also a constant over $I$, and since  $p_2= \frac{\mu_1}{(d-1) \lambda}\, p_1$, $p_2$ as well. We hence
iteratively obtain that $p_i$ is a constant for $i=1,,2,\ldots,d$. However, $\dot p_d=\- \mu_d \cdot p_d$, so that
$0=p_d=p_{d-1}=\ldots=p_1$. Finally, $p_0=0$, which is a contradiction. 
\end{IEEEproof}

\section*{Proof of Lemma.~\ref{thm:multiplier}}

\begin{IEEEproof}
The adjoint ODE system can be solved via Laplace transform. We make the replacement $\p_k(v)=p_k(T-t)$, thus considering the
backward time variable $v=T-t$. It holds $\dot \p_k(v)=-\dot p_k(t)$, so that system  \eqref{eq:adjoint} writes
\begin{eqnarray}\label{eq:adjointmod}
  &&\hskip-7mm\dot \p_0 = -\mu_0    \cdot \p_0 + \lambda d \cdot \p_1 + c_2 \beta \lambda d \nonumber\\
  && \vdots \nonumber\\
  &&\hskip-7mm\dot \p_k = -\mu_k    \cdot \p_{k-1}  + (d-k) \lambda \cdot \p_{k+1} + c_2 \beta \lambda (d-k), \nonumber \\
  && \hskip 50mm k=1,\ldots,d-1\nonumber\\
  && \vdots \nonumber\\
&&\hskip-7mm\dot \p_{d} = -\mu_d \cdot \p_{d} 
  + 2(X_d(t)-d) \cdot \1{d - X_d(t)} \p_{d+1}\nonumber\\
&&\hskip-7mm\dot \p_{d+1}=0 \nonumber
\end{eqnarray}
Let $\P_k(s)=\L\{\p_k(v)\}$ for $t=0,1,\ldots,d$ be the Laplace transform of the $k$-th variable $q_k$. The corresponding system writes
  \begin{eqnarray}\label{eq:laplace}
  &&\hskip-7mm s\P_k(s)=-\mu_k \P_k(s) + (d-k) \lambda \P_{k+1}(s)+ c_2 \beta \lambda (d-k)\frac 1s,\nonumber
  \\&& \hskip15mm \mbox{for}\; k=0,1,\ldots,d-1 \nonumber \\
 &&\hskip-7mm  s\P_d(s)=-\mu_d \P_d(s) - \gamma 
  \end{eqnarray}
  from which $\P_k(s)=\frac{\lambda (d-k) }{(s+\mu_k)} \P_{k+1}(s) + \frac{c_2 \beta\lambda (d-k)}{s(s+\mu_k)}$ is obtained. By iterative
  replacement, and by accounting for the fact that $\P_d(s)=-\frac{\gamma}{s+\mu}$, it follows
  \[
  \P_0(s)=-\frac{\gamma\lambda^d d!}{\prod_{k=0}^d (s+\mu)^k}+ \frac{c_2\beta d}s \sum_{i=0}^{d-1} \binom{d-1}{i}\frac{\lambda^{i+1}i!}{\prod_{h=0}^k (s+\mu)^h}
  \]
  In order to obtain the closed form of $p_0(t)$, auxiliary expressions of the kind $\prod_{h=0}^k (s+\mu)^h$ have to be inverted. Let us denote $f_h(t):=e^{-\mu_h t} \1{t\geq 0}$, for the sake of notation. By recalling $\L\{e^{-\mu t} \1{t\geq 0}\}=1/(s+\mu)$, it is possible to calculate
\begin{eqnarray}
  &&\hskip-7mm \L^{-1}\left \{{\prod_{h=0}^k (s+\mu)^{-h}}\right \}=f_1*\ldots*f_n= \sum_{i=0}^n \frac{e^{-\mu_i t} \1{t\geq 0}}{\prod_{\stackrel{j=0}{j\not=i}}^n{\mu_j-\mu_i}}\nonumber \\
  &&\hskip-7mm =\sum_{i=0}^n \frac{e^{-\mu_i t}\1{t\geq 0}}{\lambda^n\prod_{\stackrel{j=0}{j\not=i}}^n {i-j}}=\sum_{i=0}^n \frac{(-1)^{n-i} }{\lambda^n i!(n-i)!}f_i(t)
\end{eqnarray}
The statement follows after some algebraic manipulations of the above expression.
\end{IEEEproof}

\section*{Proof of Thm.~\ref{thm:thm1}}
\begin{IEEEproof}
From Lemma~\ref{thm:multiplier}, if $c_2=0$, it follows
\begin{equation}\label{eq:p0pure}
\dot p_0(t)=\gamma e^{-\mu(T-t)}(1-e^{-\lambda (T-t)})^{d-1} \Big ( -\mu +(\mu + \lambda d )e^{-\lambda (T-t)} \Big ) \nonumber
\end{equation}
from which it is immediate to observe that the absolute minimum over the real line is attained
at $t_{\min}=T-\frac1\lambda \log\big ( 1+\frac{d\lambda}\mu \big )$; the minimum writes
$m:=-\gamma (\frac{\mu}{\lambda d})^d/{(1+\frac{\mu}{\lambda d})^{d+\frac{\mu}{\lambda}}}$.

Switching epochs $t_s$ are determined by the instants solving $p_0(t_s)=-c_1$. First,
let observe that $p_0(T)=0>-c_1$ and $\dot p_0(T)=\lambda d$, so that
the control is indeed null in a left interval of $T$. In particular, it is possible
to identify three cases: for a given value of $\gamma>0$, there might exist either two, one or zero
switching epochs in the interior of $[0,T]$. We consider the three cases separately.

{\noindent \em Case i: single switch.} The condition for a unique switching epoch is $p_0(0)<-c_1$,
which writes  $\gamma (1 - e^{-\lambda T})^d e^{-\mu T} > c_1$, so that 
\[
\mu > \frac dT \log \Big ( \sqrt[d] {\gamma/{c_1}} (1-e^{-\lambda T})\Big ) :=\mu_0
\]
By inspection of \eqref{eq:p0pure}, due to the continuity of $p_0$, there exists switching
epoch $0 < \tsoff<T$ such that $p_0(\tsoff)=-c_1$. Because $p_0(t)$ has unimodal structure,
such switch is unique so that the corresponding optimal control
is in threshold form. Namely, $u(t)=1$ for $0\leq t<\tsoff$ and zero otherwise.

{\noindent \em Case iii: two switches.} Condition $p_0(0)>-c_1$ leads to a non-null control
if and only if $m<-c_1$. From the unimodal structure of $p_0$, and from classic
continuity arguments, there exist two real values, namely $\tson<t_{\min}<\tsoff$ where
$p_0(\tson)=-c_1=p_0(\tsoff)$, so that $u(t)=1$ for $\tson<t<\tsoff$ and zero otherwise.

{\noindent \em Case ii: no switch.} This is the case $\tson=\tsoff=0$, i.e., the optimal
control is the null one. It occurs when $p_0(0)>-c_1$ and  $m \geq -c_1$.

Finally, the explicit expression of the switching epochs is obtained by solving
equation $p_0(t)=-c_1$, which concludes the proof. 
\end{IEEEproof}

\section*{Proof of Lemma.~\ref{lem:extrema}}
\begin{IEEEproof}
It is possible to write the derivative of the multiplier $p_0(t)$ in a convenient form. For notations' sake, we
denote ${\overline p}_0(t)$ the expression of $p_0(t)$ when $c_2=0$, and $\overline t_m$ the point
(on the real line) where the minimum of ${\overline p}_0(t)$ is attained. We hence obtain
\begin{equation}\label{eq:diffp0}
\dot p_0(t)=\dot {\overline p}_0(t) - c_2 \beta \lambda d e^{-(\lambda+\mu)(T-t)}
\end{equation}
where we know that ${\overline p}_0(t_{\min})=0$, $\dot {\overline p}_0(t)<0$ for $t<t_{\min}$ and $\dot {\overline p}_0(t)>0$ for $t>t_{\min}$.

However, $p_0(T)=0$ and $\dot p_0(t)=- c_2 \beta \lambda d < 0$, so that there exists a whole left neighborhood of $T$
where $p_0(t)>0$ and decreasing. And, $\dot p_0(t) < 0 $ for $t<t_{\min}$.

By taking into account the sign of $\dot {\overline p}_0$ and the additional negative term appearing in \eqref{eq:diffp0},
it is immediate to conclude that only the following three cases are possible:
\begin{enumerate}
\item[i] $\S=\emptyset$: in this case $p_0(t)$ is strictly decreasing in $[0,T]$;
\item[ii] $\S=\{M\}$ and the maximum is attained at $0<t_M<T$: in this case $p_0(t)$ strictly increasing in $[0,t_M]$ and decreasing in $[t_M,T]$;
\item[iii] $\S=\{m,M\}$ otherwise, where $M$ is attained at $0<t_M<T$ and $m$ is attained at $0<t_m<t_M < T$ ; i.e., in this case $p_0(t)$ is decreasing in $[0,t_m]$, increasing in $[t_m,t_M]$ and then decreasing in $[t_M,T]$;
\end{enumerate}
which concludes the proof.
\end{IEEEproof}

\section*{Proof of Thm.~\ref{thm:thm_str}}
\begin{IEEEproof}
From Lemma.~\ref{lem:extrema}, the structure of the control can be analyzed exhaustively 
counting the possible switches induced by the dynamics of $p_0(t)$, similarly to what has been
done in Thm.~\ref{thm:thm1}:   
\begin{enumerate}
\item[i.] $\S=\emptyset$ implies the null control, i.e., $u \equiv 0$, i.e., $\tson=\tsoff=0$;
\item[ii.] $\S=\{M\}$ and $p_0(0)\geq -c_1$ implies the null control;
\item[iii.]  $\S=\{M\}$ and $p_0(0)< -c_1$ implies a single switch control with $\tson=0$ and $0<\tsoff<T$;  
\item[iv.] $\S=\{m,M\}$ and $p_0(0)>-c_1$ with $m>-c_1$ implies the null control;
\item[v] $\S=\{m,M\}$ with $p_0(0)<-c_1$ implies a single-switch control with $\tson=0$ and
  $0<\tsoff<T$;    
\item[vi.] $\S=\{m,M\}$ with $p_0(0)>-c_1$ with $m>-c_1$ implies a two-switch control with $\tson>0$ and
  $0<\tson<\tsoff<T$.
\end{enumerate}  
This concludes the proof, since in all cases the optimal bang-bang control is a threshold policy.
\end{IEEEproof}

\section*{Proof of Thm.~\ref{thm:mult}}
\begin{IEEEproof}
  In this proof we need to make the dependence on $\gamma$ explicit in the notation: e.g., $u_{\gamma}^*$ is
the optimal control when multiplier $\gamma$ is adopted in the relaxed objective function $J_\gamma(u)$.\\
{\noindent i.} The fact that pair $(u^*,\gamma^*)$ minimizing $J_\gamma(u)$ is unique follows from the expression $J_\gamma(u)=J(u)+\gamma(n-X_d(t))$. Let
assume by contradiction another pair $(\overline u,\overline \gamma)$ is optimal, then it must hold $J(u^*)=J(\overline u)$. However,
this implies that the two threshold policies must be identical, i.e., $u^*=\overline u$, and so also
$\gamma^*=\overline \gamma$, because of the linear dependence with multiplier $\gamma$ in \eqref{thm:multiplier}.\\
{\noindent ii.} The fact that the relaxed problem solves for the optimal solution of the original constrained minimization follows from the following argument. Let define $\U_f^n=\{\,u |\, X_d(T)=n, X_d(t)\geq d, \; t\in [0,T]\} \subset \U$, let $\gamma^*$ be the optimal multiplier and $u_*$ the optimal solution of the constrained problem. 
\begin{eqnarray}
  J(u_*) = \min_{u\in U_f^n} J(u) = \min_{u\in \U_f^n} J(u) + \gamma^* (n-X_d(T))= J_{\gamma^*}(u^*)\nonumber
  \end{eqnarray}
where the equality follows from the fact that $\gamma (n - X_d(u))=0$ over set $\U_f^n$. 
\\
{\noindent iii.}  The correctness of the bisection search is due to the fact that $J_\gamma(u^*)$ is indeed monotone in $\gamma$. In fact, costate variable 
\[
p_0^{\gamma}(t)=-\gamma \widetilde F(t)+G(t)
\]
where we have made explicit the dependence on $\gamma$ appearing in \eqref{thm:multiplier}. Now, with respect to switching epoch $\tsoff^\gamma$, let us consider
multiplier ${\gamma+\delta}$, for some $\delta>0$. Then we can write 
\[
p_0^{\gamma+\delta}(\tsoff^\gamma)=-\gamma \widetilde F(\tsoff^\gamma)+G(\tsoff^\gamma)-\delta F(\tsoff^\gamma)<0 \nonumber
\]
which implies $\tsoff^\gamma<\tsoff^{\gamma+\delta}$. Opposite holds for $\tson^\gamma$: $\tson^\gamma>\tson^{\gamma+\delta}$. From direct inspection of the
cost function, it follows $J_\gamma(u_{\gamma}^*)<J_{\gamma+\delta}(u_{\gamma+\delta}^*)$, which proves the claimed monotony argument.
\end{IEEEproof}

\end{appendix}

\end{document}